# Elemental noncentrosymmetric superconductors

Hu Zhang

*Research Center for Computational Physics of Hebei Province, Hebei Key Laboratory of High-precision Computation and Application of Quantum Field Theory, College of Physics Science and Technology, Hebei University, Baoding 071002, People's Republic of China*

E-mails: zhanghu@hbu.edu.cn

Superconductivity has been discovered in many materials including elements, cuprates and nickelates. Noncentrosymmetric superconductors get attention due to their physical properties results from spatial inversion symmetry breaking. Here we propose the existence of elemental noncentrosymmetric superconductors which needs to break spatial inversion symmetry in single element phases realized by occupations of more than one Wyckoff positions. Elemental noncentrosymmetric superconductors allow us to study the admixture of spin-singlet and spin-triplet pairing states in elements with simple crystal structures and chemical conditions. We find that four-layer La thin films coming from bulk α-La with a double hexagonal close-packed structure are elemental noncentrosymmetric superconductors since the dimension reduction breaks the spatial inversion symmetry. In the normal state, four-layer La is a polar rare-earth metal with a spontaneous polarization having a polar layer group like ferroelectrics. Hence, to be more precise, four-layer La can also be called the elemental polar superconductor in the superconductivity state. Based on this method of breaking inversion symmetry, we discovery polar rare-earth metals, polar transition metals, polar alkali metals and polar alkali earth metals. Other candidate materials are bulk polar materials formed with group 15 elements, Bi thin films with polar stacking, polar double layer plumbene and Al thin films. The question of building the periodic table of elemental noncentrosymmetric superconductors is raised.

## I. INTRODUCTION

In 1911, superconductivity was firstly discovered by Kamerlingh Onnes in element mercury having transition temperature at 4.2 K. Since then, superconductivity remains among the most studied subfields of condensed matter physics [1]. Many famous physicists, such as Albert Einstein, have considered the nature of superconductivity. An important theory of superconductivity was established by Bardeen, Cooper and Schrieffer (BCS theory) in 1957. This theory demonstrates that the superconducting state is caused by the pairing of electrons or holes (Cooper pairs) mediated by phonons. In 1986, Bednorz and Müller discovered high-temperature superconductivity in cuprates which cannot be explained by BCS theory [2]. Origins of high-temperature superconductivity are not completely known yet [3]. Since 2019, superconductivity was found in nickelate oxides including $Nd_{0.8}Sr_{0.2}NiO_2$ thin films [4], Ruddlesden-Popper nickelates $La_3Ni_2O_7$ [5] and $La_4Ni_3O_{10}$ [6,7] under high pressure. These discoveries open up a subfield of great excitement.

Another important discovery is the superconductivity in materials without spatial inversion symmetry. Aa a result of inversion symmetry breaking, the twofold spin degeneracy is lifted by spin-orbit interaction, and thus spin-singlet and spin-triplet pairings are mixed in the wave function of the Cooper pairs [8]. Such superconductors without inversion symmetry are usually called noncentrosymmetric superconductors. Magnetoelectric coupling exists in noncentrosymmetric superconductors as intriguing physical consequences of parity mixing [9]. MnSi and $CePt_3Si$ are typical noncentrosymmetric superconductors [10]. In BiPd, a complex order parameter is attributed to the admixture of singlet and triplet pairing states due to antisymmetric spin-orbit coupling [11]. A clear understanding of the unconventional nature of superconductivity in noncentrosymmetric superconductors is still missing [12].

Despite large amounts of superconductors, elemental superconductors are of both technological importance and fundamental scientific interest due to their simple crystal structures [13]. Advances in the instrumentation of physics under pressure have enabled the observation of superconductivity in some elements not previously known to superconduct. High critical temperatures ($T_c$) in elements are achieved under pressure.

Considering the existence of noncentrosymmetric superconductors, we may ask if it is possible to have elemental noncentrosymmetric superconductors, i.e. superconductivity in elements without spatial inversion symmetry. To the best of our knowledge, no such systems have been discovered theoretically and experimentally. The first step is to break spatial inversion symmetry in single element materials.

In 2018, we proposed the question of breaking spatial inversion symmetry in single element phases [14]. To this propose, elemental polar metals were predicted, which are one type of elemental metals without spatial inversion symmetry [14]. Like ferroelectrics (insulating), elemental polar metals have polar space groups and ferroelectric-like structure phase transitions. The first predicted elemental polar metals are a class of elemental metals in a distorted α-La structure with a polar space group *P6$_3$mc* in which two inequivalent Wyckoff positions are occupied by group 15 elements (P, As, Sb and Bi). We should emphasize the difference between elemental noncentrosymmetric metals and elemental polar metals. Elemental polar metals have polar space groups and thus are one special type of elemental noncentrosymmetric metals. Like ferroelectrics, there is a spontaneous polarization in a polar metal [15]. Details and histories about polar metals can be found in the review [16]. As we previously discovered, occupations of inequivalent Wyckoff positions are the key to break spatial inversion symmetry in elemental materials [14]. This rule is also important for finding out elemental noncentrosymmetric superconductors. If superconductivity can find in some elemental noncentrosymmetric metals or elemental polar metals, they will be elemental noncentrosymmetric superconductors or elemental polar superconductors. This allows us to study the admixture of spin-singlet and spin-triplet pairing states in simple elements, which might benefit from simple crystal structures and chemical conditions. Keep in mind that superconductivity was firstly found in element Hg. The discovery of elemental noncentrosymmetric superconductors would be rather interesting in the field of superconductivity.

In this work, we propose the existence of elemental noncentrosymmetric superconductors theoretically. We start from normal elemental metallic states without spatial inversion symmetry. Based on the crystal structure of α-La, elemental noncentrosymmetric (polar) metals in three-dimensional and two-dimensional systems are investigated. We transform rare-earth metals, transition metals, alkali metals, alkali earth metals, Bi, Pb and Al to polar metals. Experimental discoveries of elemental noncentrosymmetric (polar) superconductors are indicated.

## II. RESULTS AND DISCUSSION

### A. Double Hexagonal Close-Packed structures

The α-La structure is the starting point of various structures of elemental noncentrosymmetric metals in two- and three-dimension studied in this work. The crystal structure of firstly predicted elemental polar metals without spatial inversion symmetry is a distorted α-La structure [14]. The α-La structure is just the Double Hexagonal Close-Packed (DHCP) crystal structure that is not a very common crystal structure. Since lanthanum (α-La) is the first element to display this structure, DCHP may sometimes be called the La-type structure. In addition to lanthanum, cerium, praseodymium, neodymium, and promethium also have the DHCP crystal structure. In Fig. 1(a) we show the crystal structure of DHCP. The DHCP crystal structure is similar to Hexagonal Close-Packed (HCP) or Face-Centered Cubic (FCC), but it has a four-layer stacking pattern ABAC-ABAC, instead of HCP's AB-AB or FCC's ABC-ABC stacking pattern. The DHCP cell has the space group $P6_3/mmc$ (no. 194) with spatial inversion symmetry. The DHCP unit cell can be imagined as two HCP cells stacked on top of each other, but the middle layer alternates between B and C. In per primitive cell of DHCP shown in Fig. 1(a), there are four atoms occupy two inequivalent Wyckoff positions 2a (0, 0, 0) and 2c (1/3, 2/3, 1/4). The site symmetry of 2a and 2c sites are $D_{3d}$ and $D_{3h}$ respectively.

As studied in our previous work [14], when group 15 elements (phosphorus, arsenic, antimony, and bismuth) crystallize into the DHCP crystal structure they will undergo a ferroelectric structure transition from DHCP to a distorted DHCP (denoted as d-DHCP, or d-La-type) structure with a polar space group $P6_3mc$ (no. 186) without spatial inversion symmetry in which two inequivalent Wyckoff positions 2a (0, 0, $z$) and 2b (1/3, 2/3, $z$) are occupied. Such inequivalent occupied Wyckoff positions are the key to break spatial inversion symmetry in elemental materials. The primitive cell of d-DHCP is shown in Fig.1(b). The structure transition is achieved by the relative motion of atoms along the $c$ direction as indicated by arrows in the primitive cell of DHCP shown in Fig. 1(a), which results from an unstable phonon mode at the zone center (Γ point). Since there is no spatial inversion symmetry in d-DHCP, elemental metals with the d-DHCP structure are elemental noncentrosymmetric metals, or more precisely, elemental polar metals.

If we compare atomic structures shown in Fig.1(a) and Fig. 1(b) we can find that the d-DHCP structure becomes a layered structure as a result of relative motions between atoms occupied at 2a and 2c Wyckoff positions. In Fig. 1(c) we show the top view of one layer in d-DHCP, which forms the buckled honeycomb lattice. This is just the structure of monolayer blue phosphorus, silicene, germanene, stanine and plumbene in analogy with graphene. These monolayer materials have spatial inversion symmetry. Fig. 1(d) shows the Brillouin zone of the $k_z = 0$ plane, where main physics occurs.

### B. Spatial inversion symmetry breaking in multi-layer α-La

The main discovery of this work is that dimension reduction breaks spatial inversion symmetry in multi-layer α-La and thus makes it an elemental noncentrosymmetric superconductors. We choose four-layer (4L) α-La with ABAC stacking from the primitive cell of α-La in the DHCP structure shown in Fig. 1(a). To calculate physical properties of this four-layer La thin film, we use first-principles calculations based on density functional theory (DFT) [17] with the Perdew–Burke–Ernzerhoff (PBE) functional in generalized gradient approximation (GGA) [18] in the Vienna Ab Initio Simulation Package (VASP) [19-21].The fully relaxed structure is shown in Fig. 2(a) with structure parameters given. Remarkably, this two-dimensional phase has the polar layer group p3m1 (no. 69) with four Wyckoff positions 1a (0,0,$z_1$), 1a (0,0,$z_2$), 1b (1/3,2/3,$z_3$), and 1c (2/3,1/3,$z_4$) occupied. Different from centrosymmetric bulk α-La having the space group $P6_3/mmc$, the spatial inversion symmetry is breaking in α-La 4L with p3m1 symmetry. The predicted crystal lattice constant is $a = 3.72$ Å. Due to the structure reconstruction, bond lengths within the A layer ($d_1 = 3.71$ Å) and between A and B layers ($d_2 = 3.58$ Å) are different. Similarly, interval-distances between the B layer and bottom A layer ($h_2 = 2.86$ Å) or middle A layer ($h_1 = 3.11$ Å) are very different. In bulk α-La, $h_1 = h_2$. The spatial inversion symmetry in α-La 4L is broken as a result of dimension reduction from a three-dimensional DHCP structure to a two-dimensional thin film. Therefore, the crystal structure of α-La 4L displays a method to break spatial inversion symmetry in elements, which will be used to obtain various classes of polar metals in the following sections.

In Fig. 2(b), we show the phonon spectra of La 4L p3m1 along high symmetry directions in the Brillouin zone. There are no phonon modes with imaginary frequencies indicating the dynamically stability and experimental achievable of this phase. In Fig. 2(c) we show the electronic energy bands of La 4L along M-Γ-K calculated with consideration of spin-orbit coupling (SOC). The point group of $k$ at the Γ point is $C_{3v}$. Spin splitting of energy bands crossing the Fermi level $E_F$ is evident around K point due to inversion symmetry breaking. The Fermi surface of the primitive cell of La 4L is shown in Fig. 2(d). Spin splitting characters along the Γ-K direction can be found.

From above results, we demonstrate that La 4L with ABAC stacking is an elemental noncentrosymmetric metal. Or, to be more precise, La 4L is a *polar rare-earth metal*. Bulk α-La in the three-dimensional DHCP structure is a superconductor with a transition temperature $T_c = 4.8$ K [13]. Hence La 4L p3m1 found here is an elemental noncentrosymmetric superconductor or more precisely an elemental polar superconductor. Further experiments are needed to grow the La 4L thin film and measure its superconductivity carefully, especially the admixture of spin-singlet and spin-triplet pairing states. Superconductivity in bulk α-La and La 4L should be very different. In addition, layer-dependent physical properties including superconductivity in La thin films are worthy of study since thicker La films such as La 5L and La 6L are also polar metals. In experiment, ten-layer La (0001) films grown on W (110) have been reported [22]. Such thicker film is closer to bulk α-La. Thin films down to four-and five-layer need further efforts. Experimental works can also study the relation between the transition temperature and thickness of La thin films.

According our results, other polar rare-earth metals including Ce, Pr, Nb, Pm, Am, Cm, Bk, Cf 4L (or thicker films) states also exist since their bulk phases also have the DHCP structure like α-La. Bulk Am is a superconductor with a transition temperature $T_c = 1$ K [13]. Other bulk phases are not superconductors. Experimental works are needed to determine that whether superconductivity can exist in these polar rare-earth metals.

### C. Polar transition metals and superconductivity

Transition metals have rich physical properties. Motivated by above results of La, we now consider possible polar transition metals in two-dimension. At normal condition, transition metals Sc, Y, Ti, Zr, Hf, Tc, Re, Ru, Os, Co, Zn and Cd have the HCP structure with AB-AB stacking. Sc and Y are superconductors under high pressure. Tc, Re, Ru and Os are superconductors at normal pressure having transition temperature 8.2 K, 1.7 K, 0.5 K and 0.7 K respectively [13]. Like polar rare-earth metal La 4L, we study the dynamically stability of these elements in the 4L phases with ABAC stacking. Sc, Y, Tc, Re, Ru and Os are found to be dynamically stable according to their phonon spectra in the entire Brillouin zone as shown in Fig. 3. For Ru and Os, there are phonon modes with tiny imaginary frequencies around the Γ point, which often appears in calculated phonon spectra of two-dimensional materials. This is usually not the key problem since materials are grown on substrates. There are tiny imaginary frequencies around the Γ point for group 9 element Rh. Group 4 elements Ti, Zr and Hf 4L phases are less stable since each of them have one phonon mode with imaginary frequency at the M point in phonon spectra. Frequences of phonon modes at the M point are relatively small for Group 12 elements Cd indicates structural instability in a sense. The Zn 4L phase is dynamically stable. These predicted polar transition metals are worth of noting in experiments.

At normal condition, transition metals of group 10 (Ni, Pd and Pt) and 11 (Cu, Ag and Au) elements have the FCC structure with ABC-ABC stacking. In Fig. 3 we also show phonon spectra of these elements in the 4L phases. Tiny imaginary frequencies around the Γ point appear for Pt, Cu, Ag and Au. There is one unstable phonon mode for Ni. Since bulk phases of these materials are not superconductors, superconductivity in these polar states need further experimental checking. In addition, precious metals such as Pt and Pd are important catalysts. Technical applications of these *polar precious metals* will be studied elsewhere. Group 5 and 6 elements V, Nb, Ta, Cr, Mo and W (body-centered cubic for bulk, BCC) in 4L phases are not stable. Group 8 element Fe, group 9 elements Co and Ir in 4L phases are also not stable.

In Fig. 4 we show electronic structures of stable polar transition metals in 4L phases. Electronic structures of Sc and Y 4L p3m1 phases are similar to that of La 4L. Giant Rashba-type spin splitting around the Γ point can be found in Ru, Os, Rh, Pd and Pt due to their heaviness. Cu, Ag and Cu display similar properties. We also find that centrosymmetric 2L states of some of these transition metals are also stable like bismuthine [23], which will be studied elsewhere.

### D. Polar alkali metals and polar alkali earth metals

At normal condition, alkali metals (Li, Na, K, Rb and Cs) and alkali earth metals Ca, Sr and Ba have BCC and FCC structures respectively. Alkali earth metals Be and Mg have the HCP structure. Among the alkali metals only Li and Cs are found to be superconductors under pressure [13]. Under 50 GPa, Li has a $T_c$ of 20 K. The alkaline earth metal Be is a superconductor at normal pressure while Ca, Sr, and Ba are superconductors under pressure.

Similar to transition metals, we try to study the dynamically stability of alkali and alkali earth metals in the 4L phases with ABAC stacking. Our results indicate that each of Alkali metals K, Rb and Cs 4L phases have one acoustic phonon mode with large imaginary frequency around the Γ point. Li, Na and all alkali earth metals in 4L states are dynamically stable. As these materials having similar properties we select Li, Be, Mg and Sr to show phonon spectra in Fig. 5(a-d). Again, tiny imaginary frequencies around the Γ point appear for Li and Sr. Electronic structures of Li and Sr 4L phases are shown in Fig. 5(e, f). Since Li is a light element, its spin splitting is not evident. For Sr 4L, band overlaps and spin splitting appear around the K point. Superconductivity in these *polar alkali metals* and *polar alkali earth metals* need further experimental checking.

### E. Polar metals Bi, Pb and Al

Now we consider group 13, 14 and 15 elements. According to previous studies [14], elemental polar metals in the d-DHCP structure formed with group 15 elements have similar electronic structures. Theoretical crystal structure parameters of these materials can be found in Ref. [14]. We take Bi as an example to show basic physical properties of this class of three-dimensional elemental noncentrosymmetric metals. We use the crystal structure parameters ($a$ = 4.45 Å, $c$ = 8.88 Å) of Bi with the d-DHCP structure predicted in previous works. In Fig. 6(a) we plot electronic energy bands of polar phase Bi with $P6_3mc$ symmetry along high-symmetry directions of the Brillouin zone calculated with consideration of SOC. The overlap of energy bands at the Fermi level $E_F$ can be found. The point group of $k$ at the Γ point is $C_{6v}$. Due to the lack of spatial inversion symmetry, energy bands in polar phase Bi are split by SOC. Around the Γ point, Bi states form a hole pocket. An electron pocket is formed around the M point. The Fermi surface at the $k_z = 0$ plane is shown in Fig. 6(b). Six-fold rotation symmetry is clearly displayed. Other phases with $P6_3mc$ symmetry formed with group 15 elements P, As and Sb have physical properties similar to that of Bi.

From the crystal structure of d-DHCP shown in Fig. 1(b), we can also select four-layer (4L) and six-layer (6L) structures. Two phases all have the polar layer group p3m1. In Fig. 7 we show the electronic energy bands and Fermi surfaces of Bi 4L and Bi 6L. Compared with polar bulk Bi with $P6_3mc$ symmetry, they show layer-dependent properties with some similarity.

Under ordinary circumstances, bulk Bi has a rhombohedral (layered-like) structure with the space group $R-3m$ (denoted as Bi-I phase). In 2017, bulk superconductivity in pure Bi single crystals below 0.53 mK at ambient pressure was reported experimentally [24]. Such low $T_c$ may be related to its low carrier densities. Under pressure, Bi undergo phase transitions from Bi-II to Bi-V. Based on previous experimental results, Bi-II, Bi-III and Bi-V have a $T_c$ of 3.9 K, 7 K and 8 K respectively [25]. If the elemental polar metal bulk Bi $P6_3mc$ and Bi thin films with polar stacking can be grown experimentally, they may be also superconductors. In experiments, multi-layer centrosymmetric Bi thin films have been obtained. Noncentrosymmetric Bi thin films need further efforts to grow. Bulk Bi $P6_3mc$ may be obtained by layer-by-layer growth of polar stacked Bi layers.

We consider that the group 14 element Pb and group 13 element Al crystallize into the 4L p3m1 state like Bi. Both bulk Al and Pb have the FCC structure. A bi-layer structure obtained is shown in Fig. 8(a). In fact, this material is just the double layer of plumbene with polar stacking. Plumbene has been obtained in experiments [26,27]. The polar stacking means that the inversion symmetry is breaking as a result of stacking and the obtained structure has a polar layer group. The predicted crystal lattice constant is $a$ = 3.51 Å with $z_2 - z_1 = 5.8595$ Å, $z_3 - z_1 = 2.7832$ Å and $z_4 - z_1 = 8.6204$ Å. The phonon spectra of Pb p3m1 along high symmetry directions in the Brillouin zone are given in Fig. 8(b). There are no phonon modes with imaginary frequencies indicating the dynamically stability of this phase. Since plumbene has been synthesized successfully, we may expect that the polar double layer of plumbene, Pb p3m1, may also be grown experimentally in the future. In Fig. 8(c) we show the electronic energy bands of polar phase Pb p3m1 along high-symmetry directions of the Brillouin zone calculated with consideration of SOC. The point group of $k$ at the Γ point is $C_{3v}$. Spin splitting of energy bands crossing the Fermi level $E_F$ is evident due to inversion symmetry breaking. The Fermi surface of the primitive cell of Pb p3m1 is shown in Fig. 8(d). There is a more complex Fermi surface compared to that of polar bulk Bi with $P6_3mc$ symmetry and Bi thin films. Several electron pockets around the K point can be found. We find that Al in the 4L p3m1 state is also dynamically stable.

From previous experiments we know that bulk FCC Al is a superconductor with a transition temperature $T_c = 1.18$ K. Bulk FCC Pb is a strong-coupling superconductor with a transition temperature $T_c = 7.2$ K [13]. The monolayer plumbene was also predicted to be a superconductor based on the BSC theory [28]. A recent experiment in 2023 has discovered superconductivity with $T_c$ around 7 K in a buckled plumbene-Au Kagome superstructure grown by depositing Au on Pb(111) [29]. The superstructure is considered to be a monolayer Au-intercalated low-buckled plumbene (Au-plumbene) sandwiched by bottom Pb(111) and a top Au Kagome layer. From these facts it is safe to believe that superconductivity can be found in polar double layer plumbene Pb p3m1. If this is confirmed experimentally, Pb p3m1 would be an elemental noncentrosymmetric superconductor or more precisely an elemental polar superconductor. The same is true for Al p3m1.

### F. $T_c$ and bad metals

For element superconductivity, it is also well known that bad metals are often good superconductors [30]. In transition metals, $d$ electrons are relatively immobile to have any significance in the normal state conduction process but they participate in the condensation process at low temperatures that results in the superconducting state. Highly conducting, good metals (such as Cu, Ag and Au) do not become superconductors. While poorly conducting, bad metals do generally transition to the superconducting state [30]. Nb, V, Hg, Pb and Tl are poor conductors in the normal state, but they are found to superconduct with relatively high $T_c$. This is related to the fact that a strong electron–phonon interaction in metals gives a large electrical resistance above $T_c$ and a strong electron pairing at $T_c$ [30].

Polar metals are all bad metals due to the coexistence of spontaneous polarization and conductivity. Previous experiments have confirmed the large electrical resistance in polar metal $LiOsO_3$ [15]. Elemental polar superconductors with the normal states are polar metals might have relatively high $T_c$ compared to general superconducting elements. It would be interesting to check that whether the $T_c$ of polar double layer plumbene Pb p3m1 is higher than that of bulk FCC Pb (7.2 K) and monolayer plumbene. The same is true for bulk La and multi-layer La. Hence elemental polar superconductors may provide a way to find superconductors with high $T_c$. This could be beneficial for technical applications. Another open question is that whether polar precious metals, polar alkali metals and polar alkali earth metals are superconductors as a result of spontaneous polarizations.

### G. Periodic table of elemental noncentrosymmetric superconductors

Up to now, we have found superconductivity in number of elements of the periodic table. In recent years, superconductivity is discovered in many new elements under pressure. Motived by knowledge of superconducting elements, we can devote to establish the periodic table of elemental noncentrosymmetric superconductors.

Details about superconducting elements can be found in the review [13,30]. As the first element in the periodic table, hydrogen attracts much attention since metallic H is predicted to be a superconductor with critical temperatures reaching perhaps to room temperature [31]. However, no exact experimental results have been reported. Whether we can obtain noncentrosymmetric metallic H under extreme pressure is an interesting open question. At normal pressure, most of the transition metals (except Mn, Co, Ni, Cu, Ag and Au) are superconducting. Niobium holds the record for the highest $T_c$ (9.25 K) of an element at normal pressure. Metals from groups 13 to 15 (Al, Ga, In, Tl, Sn, Pb and Bi) all superconduct at normal pressure. Under pressure, O, S, Se, P, Br and I have been observed to exhibit superconductivity [13].

Start from Dmitri Mendeleev, it takes a long time to complete the periodic table of elements. Now we can build the periodic table of noncentrosymmetric superconducting elements. As the first step, we should find periodic table of elemental noncentrosymmetric metals. In Fig. 9 we show summarization of elemental noncentrosymmetric (polar) metals discovered in this work. Some of their bulk phases are superconductors. Crystal structures and transition temperatures for bulk phases are shown. Further experimental works are needed to test our predictions. In recent experiments, monolayer Au (goldene) [32] and Mo (molybdenene) [33] have been grown. Hence, it might be possible to obtain more multi-layer metals with some noncentrosymmetric stacking like La and Pb p3m1 discussed above. High pressure may also be good means to induce superconductivity. Additional works are needed to find out noncentrosymmetric metallic states for elements painted gray shown in Fig. 9.

### III. CONCLUSIONS

In conclusion, we propose the existence of elemental noncentrosymmetric superconductors. Based on the double hexagonal close-packed structure, we obtain various elemental noncentrosymmetric metals in three-dimensional and two-dimensional systems. Bulk Bi $P6_3mc$, multi-layer La and polar double layer Pb p3m1 are candidates of elemental noncentrosymmetric superconductors. Other possible materials include polar rare-earth metals, transition metals, alkali metals, alkali earth metals, and Al. The task of establishing the periodic table of elemental noncentrosymmetric superconductors is raised.

## ACKNOWLEDGMENTS

I thank M.W Ding for a stimulating discussion. This work was supported by the Advanced Talents Incubation Program of the Hebei University (Grants No. 521000981423, No. 521000981394, No. 521000981395, and No. 521000981390), the Natural Science Foundation of Hebei Province of China (Grants No. A2021201001 and No. A2021201008), the National Natural Science Foundation of China (Grants No. 12104124 and No. 12274111), the Central Guidance on Local Science and Technology Development Fund Project of Hebei Province (236Z0601G), Scientific Research and Innovation Team of Hebei University (No. IT2023B03), and the high-performance computing center of Hebei University.

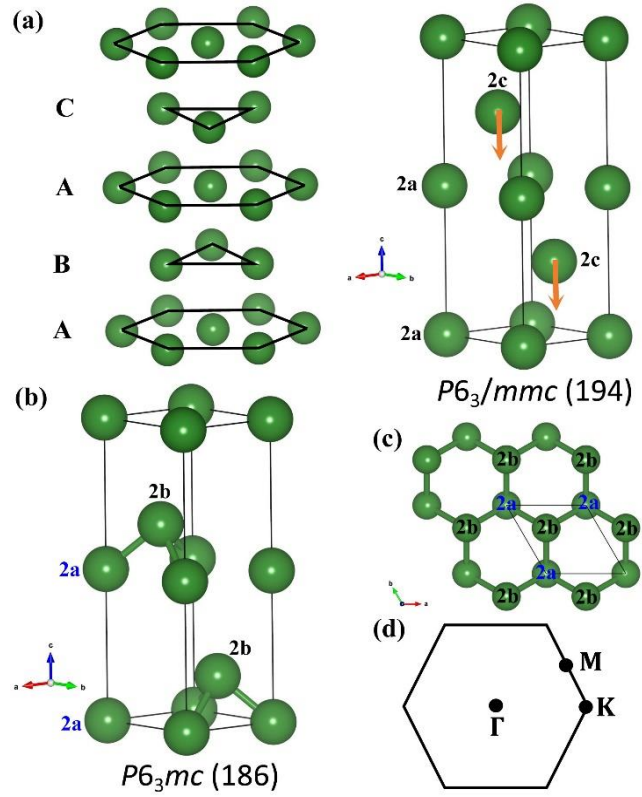

FIG. 1. The crystal structures of (a) double hexagonal close-packed (DHCP) with the $P6_3/mmc$ symmetry and (b) the distorted DHCP with $P6_3mc$ symmetry. (c) The top view of a monolayer from the distorted DHCP structure. (d) The Brillouin zone of the $k_z = 0$ plane. Wyckoff positions for each structure are given. The arrows indicate the relative atom motions between DHCP and distorted DHCP structures.

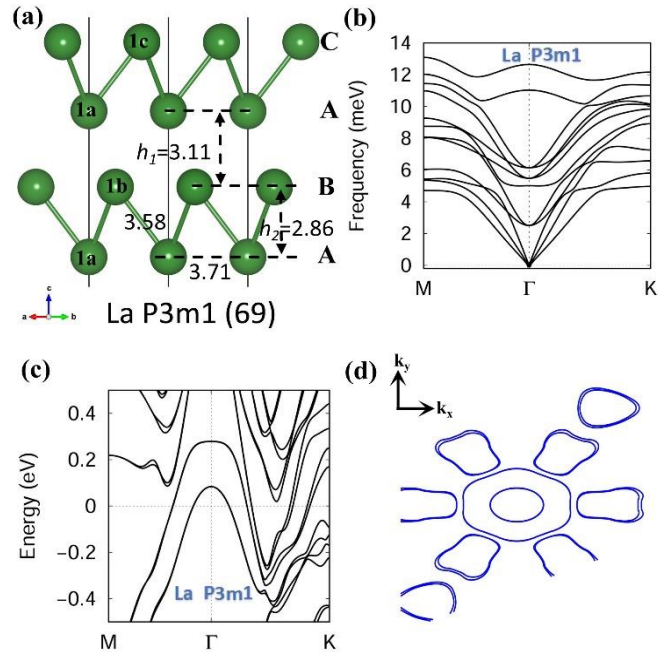

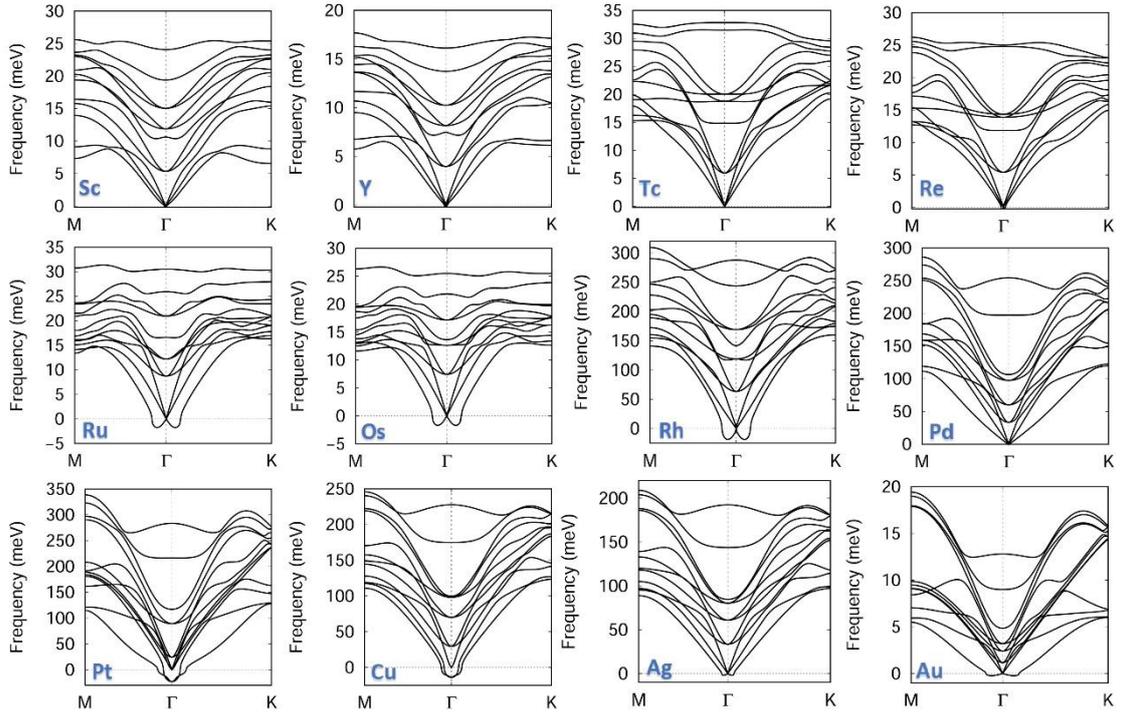

FIG. 2. (a) Side view of fully relaxed four-layer (4L) La with p3m1 symmetry obtained from the DHCP. (b) Phonon spectrum, (c) electronic band structures and (d) Fermi surfaces of La 4L.

FIG. 3. Phonon spectrum for transition metals in the four-layer state with p3m1 symmetry.

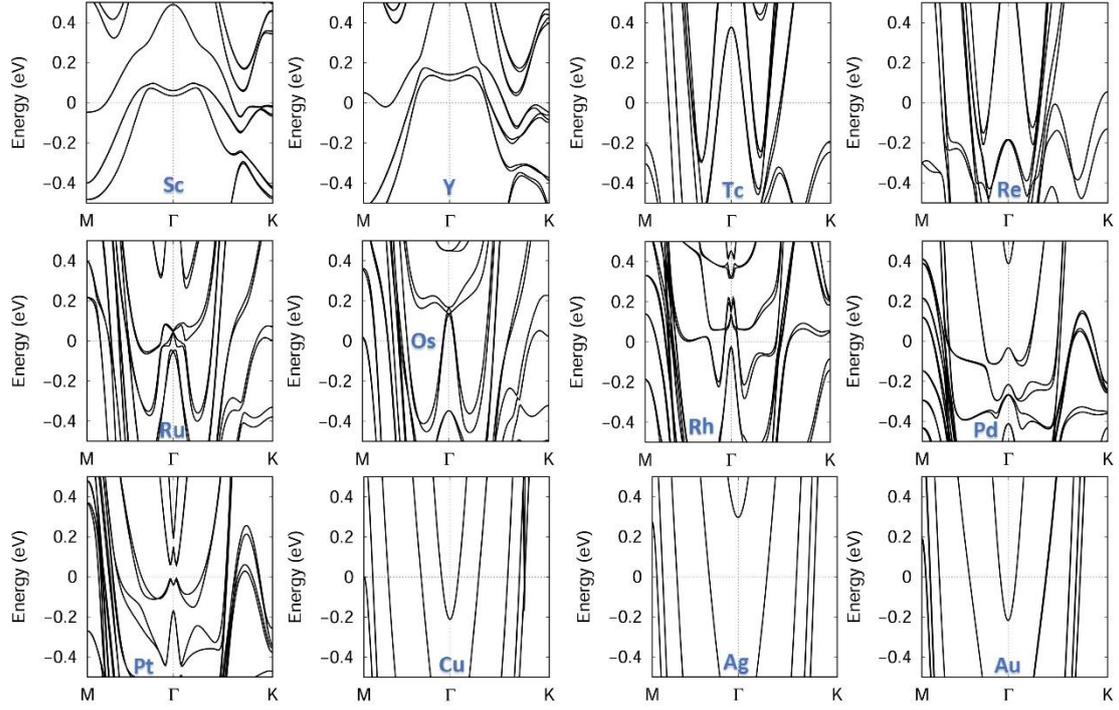

FIG. 4. Electronic band structures for transition metals in the four-layer state with p3m1 symmetry.

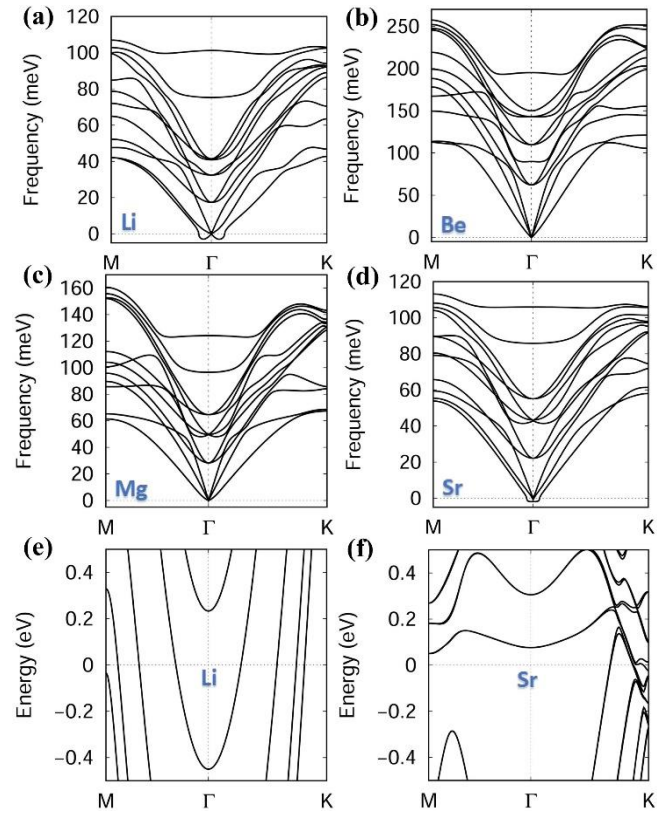

FIG. 5. (a-d) Phonon spectrum for Li, Be, Mg and Sr in the four-layer state with p3m1 symmetry. Electronic band structures for four-layer (e) Li and (f) Sr.
16

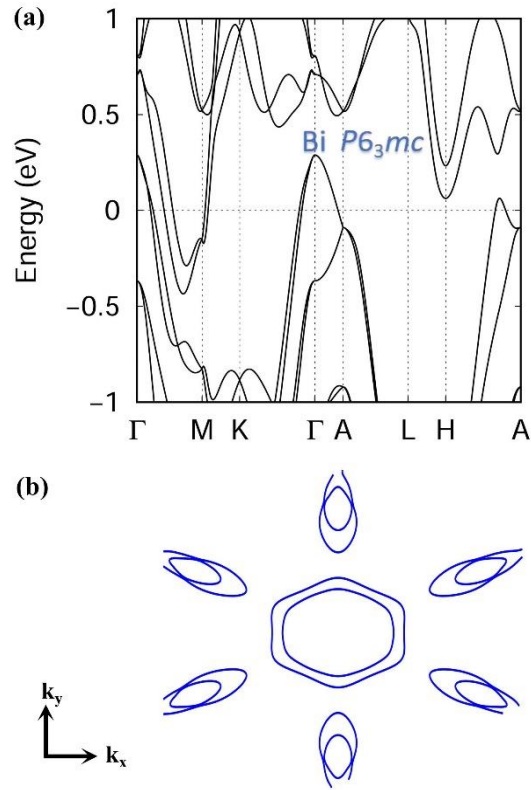

FIG. 6. (a) Electronic band structures calculated with SOC and (b) Fermi surfaces of bulk Bi with $P6_3mc$ symmetry.

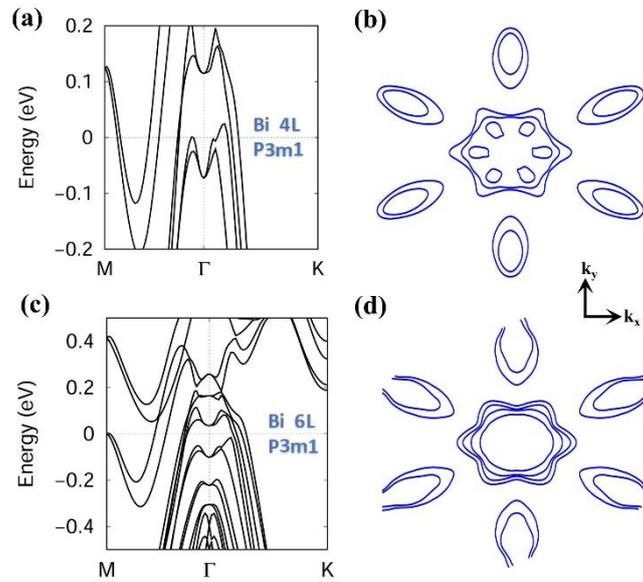

FIG. 7. Band structures and Fermi surfaces of (a,b) four-layer (4BL) Bi and (c, d) six-layer (6BL) Bi films with p3m1 symmetry.

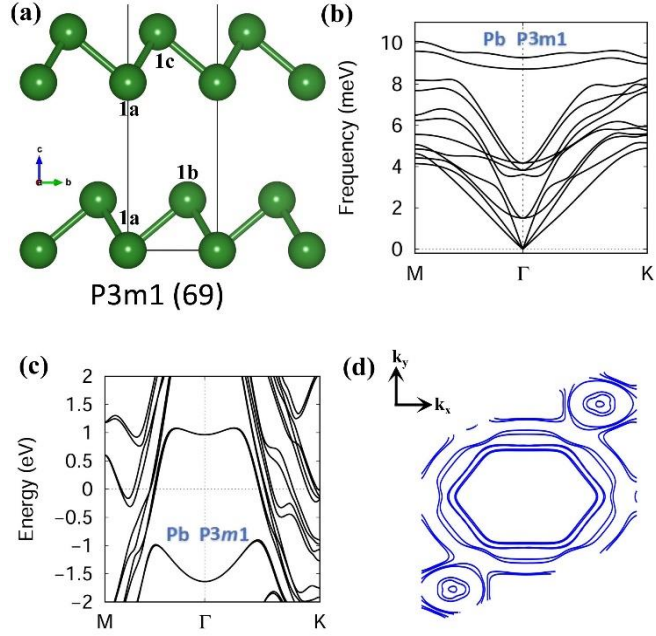

FIG. 8. (a) Side view of a polar double layer Pb with p3m1 symmetry. (b) Phonon spectrum, (c) electronic band structures and (d) Fermi surfaces of polar double layer Pb with p3m1 symmetry.

FIG. 9. Periodic table of elemental noncentrosymmetric superconductors. The crystal structures (Body-centered Cubic, Face-centered Cubic, Close packed Hexagonal, Double Close-packed Hexagonal and Rhombohedral) and transition temperatures for bulk phases are shown.

[1] P. A. Lee, N. Nagaosa, and X.-G. Wen, Doping a Mott insulator: Physics of high-temperature superconductivity, Rev. Mod. Phys. **78**, 17 (2006).

[2] J. G. Bednorz and K. A. Mller, Possible highT c superconductivity in the Ba?La?Cu?O system, Z. Phys. B: Condens. Matter **64**, 189 (1986).

[3] W. E. Pickett, Electronic structure of the high-temperature oxide superconductors, Rev. Mod. Phys. **61**, 433 (1989).

[4] D. Li, K. Lee, B. Y. Wang, M. Osada, S. Crossley, H. R. Lee, Y. Cui, Y. Hikita, and H. Y. Hwang, Superconductivity in an infinite-layer nickelate, Nature **572**, 624 (2019).

[5] H. Sun *et al.*, Signatures of superconductivity near 80 K in a nickelate under high pressure, Nature **621**, 493 (2023).

[6] M. Zhang *et al.*, Superconductivity in trilayer nickelate La4Ni3O10 under pressure, arXiv:2311.07423.

[7] Y. Zhu *et al.*, Signatures of superconductivity in trilayer La4Ni3O10 single crystals, arXiv:2311.07353.

[8] L. P. Gor'kov and E. I. Rashba, Superconducting 2D System with Lifted Spin Degeneracy: Mixed Singlet-Triplet State, Phys. Rev. Lett. **87**, 037004 (2001).

[9] V. M. Edelstein, Magnetoelectric Effect in Polar Superconductors, Phys. Rev. Lett. **75**, 2004 (1995).

[10] P. A. Frigeri, D. F. Agterberg, A. Koga, and M. Sigrist, Superconductivity without Inversion Symmetry: MnSi versusCePt3Si, Phys. Rev. Lett. **92**, 097001 (2004).

[11] L. Jiao *et al.*, Anisotropic superconductivity in noncentrosymmetric BiPd, Phys. Rev. B **89** (2014).

[12] E. Bauer and M. Sigrist, *Non-Centrosymmetric Superconductors* (Springer, Heidelberg/Germany, 2012), Vol. 847, Lecture Notes in Physics.

[13] C. Buzea and K. Robbie, Assembling the puzzle of superconducting elements: a review, Supercond. Sci. Tech. **18**, R1 (2005).

[14] H. Zhang, B. Deng, W. C. Wang, and X. Q. Shi, Parity-breaking in single-element phases: ferroelectric-like elemental polar metals, J. Phys.: Condens. Matter. **30**, 415504 (2018).

[15] Y. Shi *et al.*, A ferroelectric-like structural transition in a metal, Nat. Mater. **12**, 1024 (2013).

[16] S. Bhowal and N. A. Spaldin, Polar Metals: Principles and Prospects, Annu. Rev. Mater. Res **53**, 53 (2023).

[17] R. O. Jones, Density functional theory: Its origins, rise to prominence, and future, Rev. Mod. Phys. **87**, 897 (2015).

[18] J. P. Perdew, K. Burke, and M. Ernzerhof, Generalized Gradient Approximation Made Simple, Phys. Rev. Lett. **77**, 3865 (1996).

[19] G. Kresse and J. Furthmuller, Efficient iterative schemes for ab initio total-energy calculations using a plane-wave basis set, Phys. Rev. B **54**, 11169 (1996).

[20] M. Gajdoš, K. Hummer, G. Kresse, J. Furthmüller, and F. Bechstedt, Linear optical properties in the projector-augmented wave methodology, Phys. Rev. B **73** (2006).

[21] J. Hafner, Ab-initio simulations of materials using VASP: Density-functional theory and beyond, J Comput Chem **29**, 2044 (2008).

[22] X. Ji, Q. Chen, X. Lai, L. Huang, and S. Tan, Electronic structure of La (0001) thin films on W (110) studied by photoemission spectroscopy and first principle calculations, Sci. China Phys., Mech. **63** (2020).

[23] F. Reis, G. Li, L. Dudy, M. Bauernfeind, S. Glass, W. Hanke, R. Thomale, J. Schafer, and R. Claessen, Bismuthene on a SiC substrate: A candidate for a high-temperature quantum spin Hall material, Science **357**, 287 (2017).

[24] O. Prakash, A. Kumar, A. Thamizhavel, and S. Ramakrishnan, Evidence for bulk superconductivity in pure bismuth single crystals at ambient pressure, Science **355**, 52 (2017).

[25] Y. Li, E. Wang, X. Zhu, and H.-H. Wen, Pressure-induced superconductivity in Bi single crystals, Phys. Rev. B **95**, 024510 (2017).

[26] C. Ghosal, M. Gruschwitz, J. Koch, S. Gemming, and C. Tegenkamp, Proximity-Induced Gap Opening by Twisted Plumbene in Epitaxial Graphene, Phys. Rev. Lett. **129** (2022).

[27] G. Bihlmayer, J. Sassmannshausen, A. Kubetzka, S. Blügel, K. von Bergmann, and R. Wiesendanger, Plumbene on a Magnetic Substrate: A Combined Scanning Tunneling Microscopy and Density Functional Theory Study, Phys. Rev. Lett. **124** (2020).

[28] B. Zhang, F. Guo, M. Zhu, L. Feng, and Y. Zheng, The sensitive tunability of superconducting critical temperature in high-buckled plumbene by shifting Fermi level, Physica E **130**, 114688 (2021).

[29] W. H. Chen et al., Enhanced Superconductivity and Rashba Effect in a Buckled Plumbene-Au Kagome Superstructure, Adv. Sci. **10**, 2300845 (2023).

[30] D. R. Slocombe, V. L. Kuznetsov, W. Grochala, R. J. P. Williams, and P. P. Edwards, Superconductivity in transition metals, Philos. T. R. Soc. A: **373**, 20140476 (2015).

[31] N. W. Ashcroft, Metallic Hydrogen: A High-Temperature Superconductor?, Phys. Rev. Lett. **21**, 1748 (1968).

[32] S. Kashiwaya, Y. Shi, J. Lu, D. G. Sangiovanni, G. Greczynski, M. Magnuson, M. Andersson, J. Rosen, and L. Hultman, Synthesis of goldene comprising single-atom layer gold, Nat. Synth. (2024).

[33] T. K. Sahu et al., Microwave synthesis of molybdenene from MoS2, Nat. Nanotechnol. **18**, 1430 (2023).